\documentclass[aps,prd,twocolumn,superscriptaddress,nofootinbib]{revtex4-2}

\usepackage{graphicx}
\usepackage{amsmath}
\usepackage{amssymb}
\usepackage{bm}
\usepackage{xcolor}
\usepackage{hyperref}
\hypersetup{colorlinks=true, linkcolor=blue, citecolor=blue, urlcolor=blue}

\newcommand{\Mnu}{M_\nu}
\newcommand{\Ynu}{Y_\nu}
\newcommand{\MR}{M_R}
\newcommand{\AT}{A_T}
\newcommand{\ACP}{A_{CP}}
\newcommand{\dAT}{\delta A_T}
\newcommand{\dACP}{\delta A_{CP}}
\newcommand{\dCP}{\delta_{CP}}

\begin{document}

\title{Ultraviolet Flavor Transmission to $T$-Violating Neutrino Oscillation Observables in a Seesaw Framework with Sterile Mixing and Planck-Suppressed Corrections}

\author{Poulastya Kar}
\affiliation{Department of Physics, Kirori Mal College, University of Delhi, Delhi, India}

\author{Bipin Singh Koranga}
\affiliation{Department of Physics, Kirori Mal College, University of Delhi, Delhi, India}

\author{Vivek Nautiyal}
\affiliation{Department of Physics, Chaudhary Charan Singh University, Meerut, India}

\date{\today}

\begin{abstract}
We construct, numerically validate, and phenomenologically constrain a pipeline connecting a spontaneously CP-violating ultraviolet Type-I seesaw boundary condition to a Planck-suppressed correction of the $T$-violating neutrino oscillation asymmetry $A_T$, in a minimal $3+1$ sterile-neutrino framework. The ultraviolet Yukawa structure is reconstructed via the Casas-Ibarra parameterization, matched at the seesaw thresholds onto the effective Weinberg operator, and evolved to the electroweak scale using the established one-loop renormalization-group equations of the low-energy effective theory. The resulting renormalized operator is combined with a $3+1$ sterile sector to construct the oscillation Hamiltonian, and a Planck-suppressed correction is treated as a perturbation of this Hamiltonian using established perturbative oscillation techniques, yielding a closed-form first-order expression for $\delta A_T$. Every stage of this analytic pipeline is implemented numerically and validated independently against known analytic limits before being combined; this process identified and corrected three substantive implementation errors during development, which we report as part of the methodology rather than omit. The validated pipeline shows that the ultraviolet flavor structure is transported through renormalization-group running with high fidelity, with mixing angles and phases distorted at the per-mille level or below, even as the overall operator normalization runs by a much larger, flavor-blind factor of order 50\%, and that the Casas-Ibarra reconstruction freedom, while exactly invisible to low-energy oscillation data, is not physically inert. Subjecting the reconstructed ultraviolet texture to a four-channel phenomenological constraint suite (${\rm BR}(\mu\to e\gamma)$, ${\rm BR}(\tau\to\mu\gamma)$, ${\rm BR}(\tau\to e\gamma)$, and non-unitarity of the light PMNS matrix, together with perturbativity of the Yukawa sector) leaves $\delta A_T$ confined to a narrow range, $3\times10^{-8}$ to $3\times10^{-7}$ at the naturally expected size of the Planck-suppressed operator, several orders of magnitude below the unperturbed asymmetry itself. We further show that, by construction, $\delta A_T$ is exactly independent of the Casas-Ibarra reconstruction angles at any phenomenologically allowed point, so that the ultraviolet texture affects $\delta A_T$ only indirectly, through which heavy-neutrino spectra are permitted. A dedicated robustness check of the minimal $3+1$ sterile sector's effect on this correction finds it to be strongly texture-dependent, ranging from suppression to a factor of forty enhancement depending on the sterile mixing angles, with no universal sign or magnitude. Finally, because $A_T$ as defined here is not directly measurable by conventional accelerator-based long-baseline experiments -- which access the related but numerically distinct CP asymmetry $A_{CP}$ via the CPT theorem rather than $A_T$ itself -- we construct the corresponding antineutrino propagation Hamiltonian, verify it against the standard CPT-mediated oscillation identity, and compare the resulting Planck-induced correction to $A_{CP}$ against a sensitivity floor derived from DUNE's own published $\delta_{CP}$ resolution. We find the correction to lie roughly two to three orders of magnitude below this estimated floor. We present this as a specific, checked, and reproducible negative phenomenological result, rather than an assumed order-of-magnitude estimate or a qualitative deferral, together with an explicit accounting of the respects in which the present analysis remains incomplete.
\end{abstract}

\maketitle

\section{Introduction}
\label{sec:intro}

Neutrino oscillation experiments have firmly established that neutrinos possess non-zero masses and undergo flavor mixing, providing one of the clearest experimental indications of physics beyond the Standard Model. Precision measurements of the Pontecorvo-Maki-Nakagawa-Sakata (PMNS) matrix have significantly improved our understanding of leptonic mixing, while current and forthcoming long-baseline experiments, including DUNE and Hyper-Kamiokande, are expected to determine the leptonic CP-violating phase with unprecedented precision. Beyond the experimental determination of oscillation parameters, these measurements provide a unique opportunity to probe the flavor structure of physics operating at energy scales far beyond direct experimental reach. Understanding the origin and manifestation of leptonic CP violation has accordingly become one of the central problems in contemporary neutrino phenomenology.

From a theoretical perspective, however, the determination of low-energy oscillation parameters represents only one part of a considerably broader problem. The experimentally measured neutrino masses, mixing angles and CP-violating phases are effective low-energy quantities that arise after integrating out physical processes occurring across many orders of magnitude in energy. A question that remains comparatively underexplored is how flavor information generated within an ultraviolet theory is systematically transmitted to experimentally measurable neutrino observables. Addressing this question requires more than a successful mechanism for neutrino mass generation: it requires a consistent description of how ultraviolet flavor structures survive successive effective descriptions, threshold decouplings and quantum corrections before appearing in measurable oscillation probabilities.

The Type-I seesaw mechanism~\cite{Minkowski1977,GellMann1979,Yanagida1979,Mohapatra1980} provides one of the most compelling ultraviolet completions of neutrino mass generation, introducing heavy right-handed Majorana neutrinos whose decoupling naturally explains the observed smallness of neutrino masses. Below the seesaw scale, the resulting physics is encoded in the dimension-five Weinberg operator~\cite{Weinberg1979}. The Casas-Ibarra parameterization~\cite{CasasIbarra2001} permits the reconstruction of ultraviolet Yukawa structures consistent with observed neutrino data, while effective-field-theory threshold matching and renormalization-group evolution~\cite{Antusch2005,AntuschDrees2001} provide a systematic framework for connecting the ultraviolet and low-energy theories across multiple scales. Established perturbative treatments of neutrino propagation~\cite{Kikuchi} then enable the construction of oscillation probabilities in matter and the investigation of CP- and $T$-violating observables. These developments constitute a mature theoretical toolkit, but each component is most often applied independently or within a distinct phenomenological context.

Comparatively less attention has been devoted to following a single, physically motivated ultraviolet flavor structure through the entire sequence of effective descriptions connecting the seesaw scale to experimentally measurable $T$-violating oscillation observables. In particular, the combined treatment of spontaneous CP-violating boundary conditions, successive seesaw threshold matching, renormalization-group evolution, sterile-neutrino mixing and Planck-suppressed effective corrections within a single ultraviolet-to-low-energy pipeline has not, to our knowledge, been assembled in this specific combination.

The purpose of the present work is to construct, numerically implement, and phenomenologically constrain precisely this pipeline. We consider a Type-I seesaw framework supplemented by a minimal $3+1$ sterile-neutrino sector and a Planck-suppressed correction to the effective Weinberg operator. Beginning from spontaneous CP violation in the heavy-neutrino sector, we reconstruct the ultraviolet Yukawa structure using the Casas-Ibarra parameterization, perform successive seesaw threshold matching, evolve the resulting effective operator using the established one-loop renormalization-group equations, and construct the corresponding low-energy oscillation Hamiltonian. Using established perturbative oscillation techniques, we then derive the resulting first-order modification of the $T$-violating asymmetry $A_T$, and we implement the resulting analytic pipeline numerically, subject it to a multi-channel lepton-flavor-violation and non-unitarity constraint suite, and compare the resulting effect against an experimentally grounded sensitivity estimate for DUNE.

One terminological point is worth stating precisely at the outset, since it is easy to state too loosely. Throughout this paper, ``ultraviolet flavor information'' refers to the physical content of the effective operator $\kappa(\mu)$, equivalently of the light-neutrino mass matrix $M_\nu(\mu)$ -- not to every parameter of the ultraviolet Lagrangian individually. In particular, the Casas-Ibarra reconstruction freedom is, by construction, exactly absent from this transported information: it is verified numerically in Secs.~\ref{sec:numerics}D and~\ref{sec:numerics}G that $A_T$ depends on the ultraviolet texture only indirectly, through which heavy-neutrino spectra survive the phenomenological constraints imposed there. Section~\ref{sec:uvbc}E states this distinction in full before the numerical pipeline is developed, so that it is available to the reader from the outset rather than only in the interpretive discussion of Sec.~\ref{sec:interpretation}.

We emphasize the scope of this analysis at the outset, in order to state clearly what is and is not claimed. We do not develop new effective-field-theory matching conditions, new renormalization-group equations, or a new perturbative oscillation formalism; each of these ingredients is adopted from the existing literature without modification. The contribution of the present work is the construction, numerical validation, and phenomenological constraint of a single internally consistent chain linking these established results, together with the derivation of the resulting analytic expression for $\delta A_T$ in terms of the renormalized flavor operator, its numerical evaluation across a constrained region of parameter space, and an explicit, quantitative comparison against the CP asymmetry accessible to long-baseline experiments.

The remainder of this paper is organized as follows. Section~\ref{sec:seesaw} reviews the Type-I seesaw framework and establishes the ultraviolet neutrino mass sector employed throughout this work. Section~\ref{sec:eft} introduces the effective Weinberg operator together with the Casas-Ibarra reconstruction of the ultraviolet flavor structure. Section~\ref{sec:uvbc} specifies the ultraviolet CP boundary conditions, while Secs.~\ref{sec:threshold} and~\ref{sec:rge} implement the established threshold-matching and renormalization-group formalisms that connect the ultraviolet theory to the effective low-energy neutrino sector. Section~\ref{sec:hamiltonian} constructs the effective oscillation Hamiltonian and derives the corresponding $T$-violating observable. Section~\ref{sec:numerics} reports the numerical implementation of this pipeline: the validated transport of the UV flavor structure through renormalization-group evolution, the response of $\delta A_T$ to the Planck-suppressed correction and its regime of validity, a multi-channel phenomenological constraint scan, a robustness assessment of the sterile-neutrino sector, and an explicit comparison between the Planck-induced correction to the CP asymmetry and a DUNE-derived sensitivity floor. Section~\ref{sec:interpretation} discusses the physical interpretation of these results, Sec.~\ref{sec:conclusions} summarizes the principal conclusions, and Sec.~\ref{sec:future} states explicitly the respects in which the present analysis remains incomplete and outlines the corresponding directions for future work.

\section{Ultraviolet Flavor Framework: Type-I Seesaw}
\label{sec:seesaw}

\subsection{Ultraviolet Neutrino Mass Lagrangian}

The central objective of the present work is to investigate how flavor information generated at ultraviolet energy scales is systematically transmitted to experimentally measurable neutrino observables. The ultraviolet starting point of our analysis is the Type-I seesaw mechanism, which furnishes one of the simplest and most theoretically well-motivated extensions of the Standard Model capable of generating Majorana neutrino masses. Beyond explaining the observed smallness of neutrino masses, the Type-I seesaw introduces a non-trivial flavor structure through the neutrino Yukawa interactions and the heavy Majorana mass sector. Throughout this work these quantities are regarded as the primary carriers of ultraviolet flavor information.

The Type-I seesaw extends the Standard Model by introducing three right-handed Majorana neutrinos $N_{Ri}$, $i=1,2,3$, which are singlets under the Standard Model gauge group $SU(3)_C\times SU(2)_L\times U(1)_Y$. The leptonic Lagrangian may be written as
\begin{equation}
\mathcal{L} = -\bar L \Ynu \tilde H N_R - \frac{1}{2}\overline{N_R^c}\MR N_R + {\rm h.c.},
\label{eq:lagrangian}
\end{equation}
where $L$ denotes the Standard Model lepton doublet, $\Ynu$ is the neutrino Yukawa coupling matrix, $\MR$ is the heavy-neutrino Majorana mass matrix, and $\tilde H = i\sigma_2 H^*$ is the conjugate Higgs doublet. The first term generates Dirac neutrino masses after electroweak symmetry breaking, while the second introduces gauge-invariant Majorana masses for the heavy right-handed neutrinos.

\subsection{Electroweak Symmetry Breaking and the Full Neutrino Mass Matrix}

Following electroweak symmetry breaking, the Higgs field acquires the vacuum expectation value
\begin{equation}
\langle H\rangle = \frac{1}{\sqrt2}\begin{pmatrix}0\\v\end{pmatrix},\qquad v\simeq 246~{\rm GeV}.
\end{equation}
The Yukawa interaction generates the Dirac neutrino mass matrix
\begin{equation}
m_D = \frac{v}{\sqrt2}\Ynu.
\label{eq:mD}
\end{equation}
The complete neutrino mass matrix is therefore
\begin{equation}
\mathcal{M}_\nu = \begin{pmatrix}0 & m_D\\ m_D^T & \MR\end{pmatrix},
\label{eq:fullmassmatrix}
\end{equation}
written in the basis $(\nu_L,N_R^c)^T$. We assume throughout the conventional seesaw hierarchy $\MR\gg m_D$.

\subsection{Seesaw Reduction}

The hierarchy between the Dirac and Majorana scales enables block diagonalization of the full neutrino mass matrix. To leading order, the effective light-neutrino mass matrix is
\begin{equation}
\Mnu = -m_D^T \MR^{-1} m_D,
\label{eq:seesaw}
\end{equation}
while the heavy-neutrino masses remain approximately equal to the eigenvalues of $\MR$.

\emph{Observation.} Within the Type-I seesaw framework, the effective light-neutrino mass matrix contains no independent flavor information beyond that encoded in the ultraviolet quantities $(\Ynu,\MR)$. Consequently, any low-energy flavor observable must ultimately be traceable to these ultraviolet structures.

\subsection{Ultraviolet Flavor Information}

Equation~\eqref{eq:seesaw} is more than the familiar seesaw mass formula: within the framework developed here it identifies the ultraviolet quantities that completely determine the flavor content of the effective low-energy theory. The neutrino Yukawa matrix and the heavy Majorana mass matrix together define the ultraviolet boundary condition from which all subsequent effective descriptions are constructed. After the heavy neutrinos are integrated out, this information is encoded in the effective Weinberg operator, evolves under renormalization-group running, and ultimately determines the oscillation Hamiltonian governing experimentally measurable flavor transitions.

\subsection{Transition to the Effective Field Theory}

Neutrino oscillation experiments operate at energies far below the seesaw scale, where the heavy Majorana neutrinos no longer appear as explicit degrees of freedom. The appropriate description is an effective field theory in which the heavy states have been integrated out and their flavor information is encoded in the effective Weinberg operator. The next section develops this effective description and introduces the Casas-Ibarra reconstruction of the ultraviolet flavor structure.

\section{Effective Field Theory Bridge: From the Ultraviolet Theory to the Low-Energy Flavor Sector}
\label{sec:eft}

\subsection{Integrating Out the Heavy Neutrino Sector}

At leading order in the inverse heavy-neutrino mass, integrating out the heavy states generates the unique dimension-five operator consistent with the Standard Model gauge symmetries, the Weinberg operator,
\begin{equation}
\mathcal{L}^{(5)}_{\rm eff} = \frac{1}{2}\kappa_{\alpha\beta}(L_\alpha^c\tilde H^*)(L_\beta\tilde H) + {\rm h.c.},
\label{eq:weinberg}
\end{equation}
where $\kappa$ is the effective Wilson coefficient matrix and Greek indices label lepton flavors.

\subsection{Matching the Type-I Seesaw onto the Weinberg Operator}

The effective coupling matrix is obtained through the standard tree-level matching condition at the seesaw threshold. Combining Eq.~\eqref{eq:seesaw} with Eq.~\eqref{eq:mD},
\begin{equation}
\kappa = -\Ynu^T \MR^{-1}\Ynu.
\label{eq:matching}
\end{equation}
This matching relation transfers the flavor structure already contained in $(\Ynu,\MR)$ into the Wilson coefficient of the effective operator without generating new flavor information.

\subsection{Reconstruction of the Ultraviolet Flavor Structure}

While the effective neutrino mass matrix is constrained experimentally through oscillation measurements, the corresponding ultraviolet Yukawa couplings are not directly accessible. For this purpose we employ the Casas-Ibarra parameterization~\cite{CasasIbarra2001}, which provides the most general neutrino Yukawa matrix consistent with the observed neutrino masses and leptonic mixing. Diagonalizing the effective neutrino mass matrix,
\begin{equation}
U^T\Mnu U = D_\nu \equiv {\rm diag}(m_1,m_2,m_3),
\label{eq:diag}
\end{equation}
the corresponding Yukawa matrix may be written as
\begin{equation}
\Ynu = \frac{i\sqrt2}{v}\MR^{1/2}\,R\,D_\nu^{1/2}\,U^\dagger,
\label{eq:casasibarra}
\end{equation}
where $R$ is a complex orthogonal matrix satisfying $R^TR=\mathbb{1}$. We note explicitly, as established during the numerical implementation of Sec.~\ref{sec:numerics}, that the overall factor of $i$ in Eq.~\eqref{eq:casasibarra} is required for self-consistency with the seesaw relation Eq.~\eqref{eq:seesaw}; this factor is standard in the original Casas-Ibarra construction and in downstream implementations, and its presence here is confirmed by an explicit round-trip reconstruction test (Sec.~\ref{sec:numerics}A). The Casas-Ibarra parameterization is used throughout this work solely as a reconstruction tool: it allows an ultraviolet flavor structure consistent with current oscillation data to be reconstructed without additional assumptions about the heavy-neutrino dynamics. We make no claim of originality regarding Eqs.~\eqref{eq:diag}--\eqref{eq:casasibarra}, which are adopted directly from Ref.~\cite{CasasIbarra2001}.

\subsection{The Weinberg Operator as the Effective Flavor Carrier}

The matching relation Eq.~\eqref{eq:matching} together with the Casas-Ibarra reconstruction Eq.~\eqref{eq:casasibarra} establishes a complete correspondence between the ultraviolet and effective descriptions,
\begin{equation}
(\Ynu,\MR)\;\longrightarrow\;\kappa\;\longrightarrow\;\Mnu.
\label{eq:chain1}
\end{equation}
Once the heavy-neutrino degrees of freedom are integrated out, every subsequent stage of the analysis depends exclusively on the evolution of this effective operator.

\subsection{Transition to the Ultraviolet Boundary Conditions}

The remaining question is not how the effective theory is obtained, but what flavor structure is specified at the ultraviolet scale before the effective evolution begins. Section~\ref{sec:uvbc} introduces the ultraviolet CP boundary conditions adopted in this analysis.

\section{Ultraviolet Boundary Conditions and Flavor Structure}
\label{sec:uvbc}

\subsection{Ultraviolet CP Symmetry}

We assume that the underlying ultraviolet Lagrangian is initially invariant under the combined charge-conjugation and parity (CP) transformation. Prior to spontaneous symmetry breaking, the neutrino Yukawa couplings and the heavy Majorana mass matrix may therefore be chosen to be real,
\begin{equation}
\Ynu = \Ynu^*,\qquad \MR = \MR^*,
\end{equation}
so that no explicit CP-violating parameters are present in the fundamental Lagrangian.

\subsection{Spontaneous Generation of Flavor Phases}

To generate a physically non-trivial flavor structure, we assume that CP symmetry is broken spontaneously at the ultraviolet scale through the vacuum expectation value of a scalar field,
\begin{equation}
\langle\Phi\rangle = v_\phi\, e^{i\alpha},
\end{equation}
where the phase $\alpha$ characterizes the spontaneous breaking of CP symmetry. The purpose of this work is not to construct a microscopic model of spontaneous CP violation; mechanisms of this kind are well established in the flavor-model literature~\cite{King2015,Nishi2023}. Instead, the spontaneously generated flavor phases are treated as physically motivated ultraviolet boundary conditions from which the effective flavor evolution begins. In the numerical implementation of Sec.~\ref{sec:numerics}, this freedom is represented concretely by a benchmark choice of the complex Casas-Ibarra angle $R$.

\subsection{Boundary Conditions for the Effective Operator}

Using the matching relation Eq.~\eqref{eq:matching}, the ultraviolet flavor structure is immediately transferred to the effective Weinberg operator, $\kappa=\kappa(\Ynu,\MR)$.

\subsection{Physical Flavor Degrees of Freedom}

Although spontaneous symmetry breaking introduces complex parameters into the Yukawa sector, not every complex phase corresponds to a physical observable. The present analysis is formulated in terms of basis-independent flavor information contained in the effective neutrino mass operator, consistent with the basis-independent, flavor-invariant treatment of seesaw CP violation developed in Ref.~\cite{Invariants}.

\subsection{Physical Interpretation}

Sections~\ref{sec:seesaw}--\ref{sec:uvbc} establish the complete ultraviolet description used throughout this analysis, summarized as
\begin{equation}
(\Ynu,\MR)\;\xrightarrow{\text{seesaw}}\;\kappa\;\xrightarrow{\text{Casas-Ibarra}}\;\Mnu.
\label{eq:chain2}
\end{equation}

We use the phrase ``ultraviolet flavor information'' throughout this paper to refer specifically to the physical content of the effective operator $\kappa(\mu)$ -- equivalently, of the light-neutrino mass matrix $\Mnu(\mu)$ -- rather than to every ultraviolet parameter individually. This distinction matters concretely: the Casas-Ibarra reconstruction of Sec.~\ref{sec:eft}C guarantees that $\Mnu(\MR)$, and therefore $\kappa(\MR)$, depends only on the low-energy data (light masses and the PMNS matrix, including $\dCP$) and on the heavy spectrum $\MR$, and is exactly independent of the complex-orthogonal matrix $R$ -- a property verified numerically in Sec.~\ref{sec:numerics}D and shown there to hold to floating-point precision. The Casas-Ibarra angles therefore parameterize a genuine ultraviolet ambiguity in $(\Ynu,\MR)$ individually, but not a piece of information that is ``transported'' to the $T$-violating observable $\AT$ in the sense used throughout this paper; Sec.~\ref{sec:numerics}G confirms this explicitly, finding that $\AT$ depends on the ultraviolet texture only indirectly, through which $(\MR,R)$ pairs survive the phenomenological constraints of that section. We flag this here, before the pipeline is developed further, precisely so that the later, more precise statement in Sec.~\ref{sec:interpretation} is not the first place a reader encounters it.

We also note, for the same reason, the limited and specific role played by the spontaneous CP-violating mechanism introduced in Sec.~\ref{sec:uvbc}B. That mechanism motivates the presence of a nonzero low-energy CP phase $\dCP$ in the input PMNS matrix used throughout Secs.~\ref{sec:threshold}--\ref{sec:interpretation}; it does not, and is not used to, constrain or motivate the complex structure of the Planck-suppressed Wilson coefficient introduced in Sec.~\ref{sec:numerics}A, whose six complex basis coefficients are treated there as a generic, phenomenologically unconstrained input. Section~\ref{sec:uvbc} should accordingly be read as background motivation for taking $\dCP\neq0$ in the ultraviolet boundary condition -- addressing the objection that a complex low-energy phase might otherwise appear to be assumed rather than derived from some named mechanism -- and not as a mechanism from which any numerical result of Sec.~\ref{sec:numerics} is derived. Connecting the origin of the Planck operator's complex structure to the same or a related spontaneous-symmetry-breaking sector is a natural but nontrivial extension that we do not attempt here (see Sec.~\ref{sec:future}).

\subsection{Transition to Threshold Matching}

Having specified both the ultraviolet flavor framework and the corresponding boundary conditions, we next determine how the effective operator $\kappa$, defined in the precise sense above, is transmitted across successive effective descriptions as the heavy Majorana neutrinos decouple.

\section{Encoding Ultraviolet Flavor Information into the Effective Theory}
\label{sec:threshold}

\subsection{Sequential Decoupling of the Heavy-Neutrino Sector}

Assuming the conventional hierarchical spectrum $M_1<M_2<M_3$, the effective theory evolves through three successive threshold transitions,
\begin{equation}
\mathcal{L}_{\rm UV}\to\mathcal{L}^{(3)}_{\rm EFT}\to\mathcal{L}^{(2)}_{\rm EFT}\to\mathcal{L}^{(1)}_{\rm EFT},
\end{equation}
where each effective theory is valid only within the corresponding energy interval. Throughout this work we employ the standard sequential decoupling procedure developed in the neutrino effective-field-theory literature~\cite{Antusch2005}.

\subsection{Threshold Matching of the Weinberg Operator}

Integrating out a heavy Majorana neutrino generates an effective contribution to the dimension-five Weinberg operator. At each threshold the Wilson coefficient is fixed by the standard tree-level matching condition
\begin{equation}
\kappa^{(i)} = -\Ynu^{(i)T}M_i^{-1}\Ynu^{(i)},
\end{equation}
where $\Ynu^{(i)}$ is the Yukawa matrix immediately above the corresponding threshold. Successive decoupling produces a sequence of effective Wilson coefficients, $\kappa(M_3)\to\kappa(M_2)\to\kappa(M_1)$.

\subsection{Encoding Ultraviolet Flavor Information}

Above the seesaw scale, flavor information resides explicitly in $\Ynu$ and $\MR$. Below threshold, these quantities no longer appear as dynamical variables, yet their complete physical content survives through the effective Wilson coefficient, $(\Ynu,\MR)\to\kappa(\mu=M_i)$.

\subsection{Planck-Suppressed Contributions to the Effective Operator}

While the conventional Type-I seesaw determines the dominant contribution to the effective Weinberg operator, additional ultraviolet physics may induce subleading corrections suppressed by the Planck scale,
\begin{equation}
\kappa = \kappa_{\rm SS} + \delta\kappa_{\rm Pl},\qquad \|\delta\kappa_{\rm Pl}\|\ll\|\kappa_{\rm SS}\|.
\label{eq:planckcorrection}
\end{equation}
Rather than introducing an independent modification of the oscillation Hamiltonian, the Planck-scale contribution is treated as a perturbation of the effective operator itself. The detailed flavor structure of $\delta\kappa_{\rm Pl}$ is specified in the numerical implementation of Sec.~\ref{sec:numerics}, where a basis-complete parameterization is introduced and its impact on $T$-violating observables is evaluated quantitatively.

\subsection{Transition to Renormalization-Group Evolution}

The effective Weinberg operator obtained after threshold matching remains scale dependent and continues to evolve under quantum corrections between the seesaw scale and experimentally accessible energies. Section~\ref{sec:rge} employs the established one-loop renormalization-group equations of the effective theory to transport the encoded flavor information across energy scales.

\section{Transport of Ultraviolet Flavor Information: Renormalization-Group Evolution}
\label{sec:rge}

\subsection{Renormalization-Group Evolution Below the Seesaw Scale}

Following the successive threshold matching discussed in Sec.~\ref{sec:threshold}, the heavy Majorana neutrinos have been integrated out and the ultraviolet flavor structure is completely encoded in the Wilson coefficient of the effective Weinberg operator. Throughout this work we describe the subsequent evolution using the standard one-loop renormalization-group equations of the effective neutrino field theory~\cite{Antusch2005,AntuschDrees2001}.

\subsection{Evolution of the Effective Weinberg Operator}

The scale dependence of the effective theory is governed by the renormalization-group equation
\begin{equation}
16\pi^2\frac{d\kappa}{dt} = \beta_\kappa,\qquad t=\ln\mu,
\label{eq:rge}
\end{equation}
where $\beta_\kappa$ is the corresponding one-loop beta function. We employ the explicit form derived by Antusch, Drees, Kersten, Lindner, and Ratz~\cite{AntuschDrees2001}, which corrects an earlier coefficient in the literature and reads, in the flavor basis where the charged-lepton Yukawa matrix $Y_e$ is diagonal,
\begin{equation}
16\pi^2\beta_\kappa = -\frac{3}{2}\left[\kappa(Y_e^\dagger Y_e) + (Y_e^\dagger Y_e)^T\kappa\right] + C\kappa,
\label{eq:betakappa}
\end{equation}
where $C$ collects the flavor-blind gauge-coupling, Higgs-quartic, and quark/lepton-trace contributions that rescale $\kappa$ proportionally to the identity in flavor space. This form is employed without modification throughout this work; its numerical implementation and validation are described in Sec.~\ref{sec:numerics}.

\subsection{Evolution of the Effective Neutrino Mass Matrix}

At every renormalization scale, the effective neutrino mass matrix is reconstructed from the Wilson coefficient through
\begin{equation}
\Mnu(\mu) = \frac{v^2}{2}\kappa(\mu).
\label{eq:mnurun}
\end{equation}
Diagonalization of the running mass matrix, $U^T(\mu)\Mnu(\mu)U(\mu)=D_\nu(\mu)$, determines the running neutrino masses, mixing angles and CP-violating phases.

\subsection{Transport of Flavor Information}

The sequence $(\Ynu,\MR)\to\kappa(M_i)\to\kappa(M_Z)$ summarizes the complete ultraviolet-to-low-energy evolution of the effective flavor operator, and is confirmed numerically in Sec.~\ref{sec:numerics}B--C, where the relative size of the flavor-blind and flavor-dependent parts of this running are evaluated explicitly.

\subsection{Evolution of Planck-Suppressed Contributions}

Since both $\kappa_{\rm SS}$ and $\delta\kappa_{\rm Pl}$ are components of the same effective operator, they evolve under the same renormalization-group equation, $16\pi^2\,d(\kappa_{\rm SS}+\delta\kappa_{\rm Pl})/dt = \beta_\kappa$. We introduce no independent renormalization-group equations for the Planck-scale contribution.

\subsection{Low-Energy Flavor Operator}

After renormalization-group evolution, the Wilson coefficient evaluated near the electroweak scale, $\kappa(M_Z)$, contains the complete flavor information required for the subsequent oscillation analysis:
\begin{equation}
(\Ynu,\MR)\to\kappa(M_i)\to\kappa(M_Z)\to\Mnu(M_Z).
\end{equation}

\subsection{Transition to Oscillation Phenomenology}

The renormalized effective mass matrix obtained in this section provides the complete input required for constructing the low-energy neutrino propagation Hamiltonian, developed next in Sec.~\ref{sec:hamiltonian}.

\section{Mapping Ultraviolet Flavor Information to $T$-Violating Observables}
\label{sec:hamiltonian}

\subsection{Construction of the Low-Energy Oscillation Hamiltonian}

The renormalization-group evolution of Sec.~\ref{sec:rge} provides the effective neutrino mass operator at the electroweak scale, $\Mnu(M_Z)$. For relativistic neutrinos propagating in vacuum,
\begin{equation}
H_{\rm vac} = \frac{1}{2E}U\,\Mnu^2(M_Z)\,U^\dagger,
\label{eq:hvac}
\end{equation}
where $U$ is the PMNS matrix obtained by diagonalizing the renormalized neutrino mass matrix.

\subsection{Matter Effects and Sterile-Neutrino Mixing}

For neutrinos propagating through ordinary matter, coherent forward scattering introduces the standard charged-current matter potential,
\begin{equation}
V_{CC} = \sqrt2\,G_F N_e.
\end{equation}
Extending the framework to a minimal $3+1$ scenario, the flavor basis is enlarged to include one sterile neutrino, $(\nu_e,\nu_\mu,\nu_\tau,\nu_s)^T$, and the mixing matrix becomes $U\to U_{4\times4}$. The complete propagation Hamiltonian is
\begin{equation}
H_{\rm eff} = \frac{1}{2E}U_{4\times4}\Mnu^2(M_Z)U_{4\times4}^\dagger + V_{\rm matter}.
\label{eq:heff4}
\end{equation}
For antineutrinos, the same construction applies with the vacuum piece charge-conjugated, $H_{\rm vac}\to H_{\rm vac}^*$, and the matter potential sign-reversed, $V_{CC}\to-V_{CC}$; this prescription is verified numerically in Sec.~\ref{sec:numerics}H against the standard analytic identity relating neutrino and antineutrino oscillation probabilities under $\dCP\to-\dCP$.

\subsection{Planck-Suppressed Perturbation of the Effective Flavor Operator}

The renormalized effective neutrino mass matrix contains both the seesaw contribution and the small Planck-suppressed correction,
\begin{equation}
\Mnu = M_{\rm SS} + \delta M_{\rm Pl},\qquad \delta M_{\rm Pl} = \frac{v^2}{2}\delta\kappa_{\rm Pl},
\end{equation}
with $\|\delta M_{\rm Pl}\|\ll\|M_{\rm SS}\|$. To first order, the squared mass matrix is $\Mnu^2 = M_{\rm SS}^2 + M_{\rm SS}\delta M_{\rm Pl} + \delta M_{\rm Pl}M_{\rm SS} + \mathcal{O}(\delta M_{\rm Pl}^2)$, so that the oscillation Hamiltonian decomposes naturally as
\begin{equation}
H_{\rm eff} = H_0 + \delta H,
\label{eq:hsplit}
\end{equation}
where
\begin{align}
H_0 &= \frac{1}{2E}U M_{\rm SS}^2 U^\dagger + V_{\rm matter},\\
\delta H &= \frac{1}{2E}U\left[M_{\rm SS}\delta M_{\rm Pl}+\delta M_{\rm Pl}M_{\rm SS}\right]U^\dagger.
\label{eq:deltaH}
\end{align}
Unlike conventional perturbative treatments, in which $\delta H$ is introduced phenomenologically, the perturbation here follows directly from the renormalized effective flavor operator constructed through the ultraviolet-to-infrared evolution of Secs.~\ref{sec:seesaw}--\ref{sec:rge}.

\subsection{Perturbative Evolution of the Oscillation Operator}

Having obtained the perturbation induced by the effective flavor operator, we use the established perturbative oscillation formalism~\cite{Kikuchi} to determine its effect on neutrino propagation. The evolution operator satisfies $S(L)=e^{-iH_{\rm eff}L}$, which to first order in $\delta H$ becomes
\begin{equation}
S(L) = S_0(L) - i\int_0^L S_0(L-x)\,\delta H\,S_0(x)\,dx + \mathcal{O}(\delta H^2),
\label{eq:perturbativeS}
\end{equation}
where $S_0(L)=e^{-iH_0L}$. This expansion is numerically validated in Sec.~\ref{sec:numerics}E against the exact, non-perturbative evolution operator, which establishes its explicit regime of validity rather than assuming one.

\subsection{$T$-Violating Oscillation Observable}

The oscillation probability is $P_{\alpha\beta}=|S_{\beta\alpha}|^2$, and the corresponding $T$-violating asymmetry is
\begin{equation}
\AT = P(\nu_\mu\to\nu_e) - P(\nu_e\to\nu_\mu).
\label{eq:AT}
\end{equation}
Using the first-order expansion $S=S_0+\delta S$, the asymmetry becomes $\AT = A_T^{(0)}+\dAT$, where
\begin{equation}
\dAT = 2\,{\rm Re}\!\left[S_{0,e\mu}^*\,\delta S_{e\mu} - S_{0,\mu e}^*\,\delta S_{\mu e}\right].
\label{eq:dAT}
\end{equation}
Equation~\eqref{eq:dAT} establishes the direct, explicit connection between the Planck-suppressed correction to the effective flavor operator and an experimentally measurable $T$-violating oscillation observable. It is the central analytic result of the present work, and its numerical evaluation -- including an explicit treatment of the sense in which $\AT$ differs from the CP asymmetry actually accessible to conventional long-baseline experiments -- is the subject of Sec.~\ref{sec:numerics}.

\subsection{Physical Interpretation}

The derivation above completes the ultraviolet-to-observable mapping developed throughout this work, summarized as
\begin{equation}
(\Ynu,\MR)\to\kappa\to\kappa(M_Z)\to\Mnu(M_Z)\to H_{\rm eff}\to\AT.
\label{eq:fullchain}
\end{equation}

\subsection{Transition to Numerical Phenomenology}

The analytical framework developed in this section establishes the connection between ultraviolet flavor dynamics and a measurable $T$-violating observable. What remains is to determine the quantitative size of this effect for physically relevant regions of parameter space, to assess whether that region survives existing phenomenological constraints, and to compare the result against the sensitivity of next-generation experiments. Section~\ref{sec:numerics} carries out this numerical program.

\section{Numerical Implementation and Results}
\label{sec:numerics}

\subsection{Overview}

Section~\ref{sec:hamiltonian} derived the first-order modification of the $T$-violating asymmetry, $\dAT$, induced by a Planck-suppressed correction to the effective Weinberg operator, Eq.~\eqref{eq:dAT}. We now report the results of the numerical implementation of the full pipeline summarized in Fig.~\ref{fig:1}, from the ultraviolet seesaw boundary condition through threshold matching, renormalization-group evolution, and into $\dAT$ itself. Every stage of this implementation has been validated independently against known limits (analytic RG-off and $Y_e\to0$ limits, the standard two-flavor vacuum oscillation formula, solver-tolerance convergence) before being combined; these validations are documented in the accompanying code and are not repeated in full here, but their outcomes are quoted where they bear directly on the reliability of a given figure.

Throughout this section we work with a fixed illustrative ultraviolet texture unless stated otherwise: a hierarchical heavy spectrum $\MR = (0.1,0.3,1)\times M_{R,3}$, light-neutrino parameters fixed to their NuFIT 6.0~\cite{NuFIT6} best-fit values (normal ordering), and a Casas-Ibarra rotation $R$ with a single nonzero complex angle $z_{12}=0.3+i\,{\rm Im}(z_{12})$, used as a concrete (not derived) stand-in for the flavor structure generated by the spontaneous ultraviolet CP breaking of Sec.~\ref{sec:uvbc}. We emphasize that this is one representative slice of a much larger parameter space; where a result depends sensitively on this choice, we say so explicitly rather than presenting it as texture-independent.

\subsection{UV-to-IR Transport of the Effective Operator}

\begin{figure}[t]
\centering
\includegraphics[width=\columnwidth]{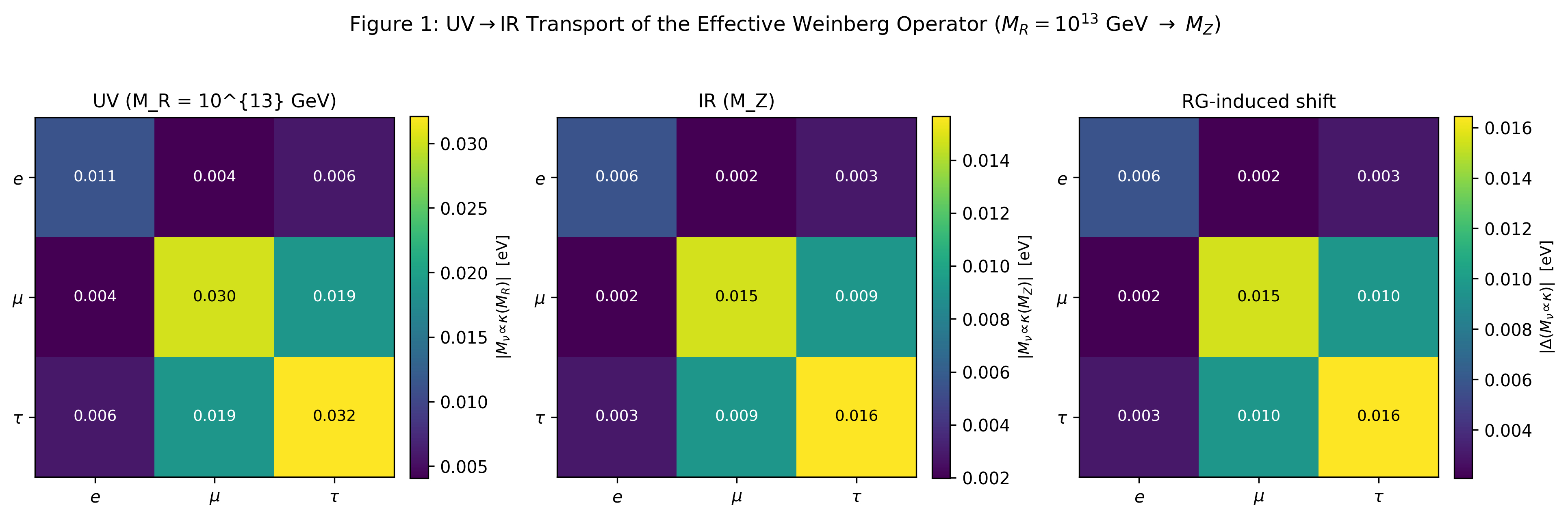}
\caption{Ultraviolet-to-infrared transport of the effective Weinberg operator for the benchmark texture of Sec.~\ref{sec:numerics}A. Each panel shows $|v^2\kappa/2|$ (equivalent mass-matrix magnitude, in eV) in the charged-lepton flavor basis. Left: at the seesaw scale $\MR=10^{13}$~GeV. Center: after one-loop renormalization-group evolution to $M_Z$. Right: their difference. The overall normalization decreases by $\|\Delta\kappa\|/\|\kappa(\MR)\|\simeq0.51$ between the two scales; Fig.~\ref{fig:2} shows this is almost entirely a flavor-blind rescaling.}
\label{fig:1}
\end{figure}

Figure~\ref{fig:1} shows the effective Weinberg operator, expressed as the equivalent mass-matrix magnitude $|v^2\kappa/2|$, at the seesaw scale $\MR=10^{13}$~GeV and at $M_Z$, together with their difference, for the benchmark texture above. The renormalization-group evolution of Sec.~\ref{sec:rge} produces a relative change $\|\Delta\kappa\|/\|\kappa(\MR)\|\simeq0.51$ between these two scales: roughly half of the operator's overall normalization is washed out by running. This is a substantial effect, but -- as Fig.~\ref{fig:2} clarifies -- it is almost entirely a flavor-blind rescaling, not a distortion of the flavor structure itself.

\subsection{RG Trajectories and the Flavor-Dependent Running}

\begin{figure}[t]
\centering
\includegraphics[width=\columnwidth]{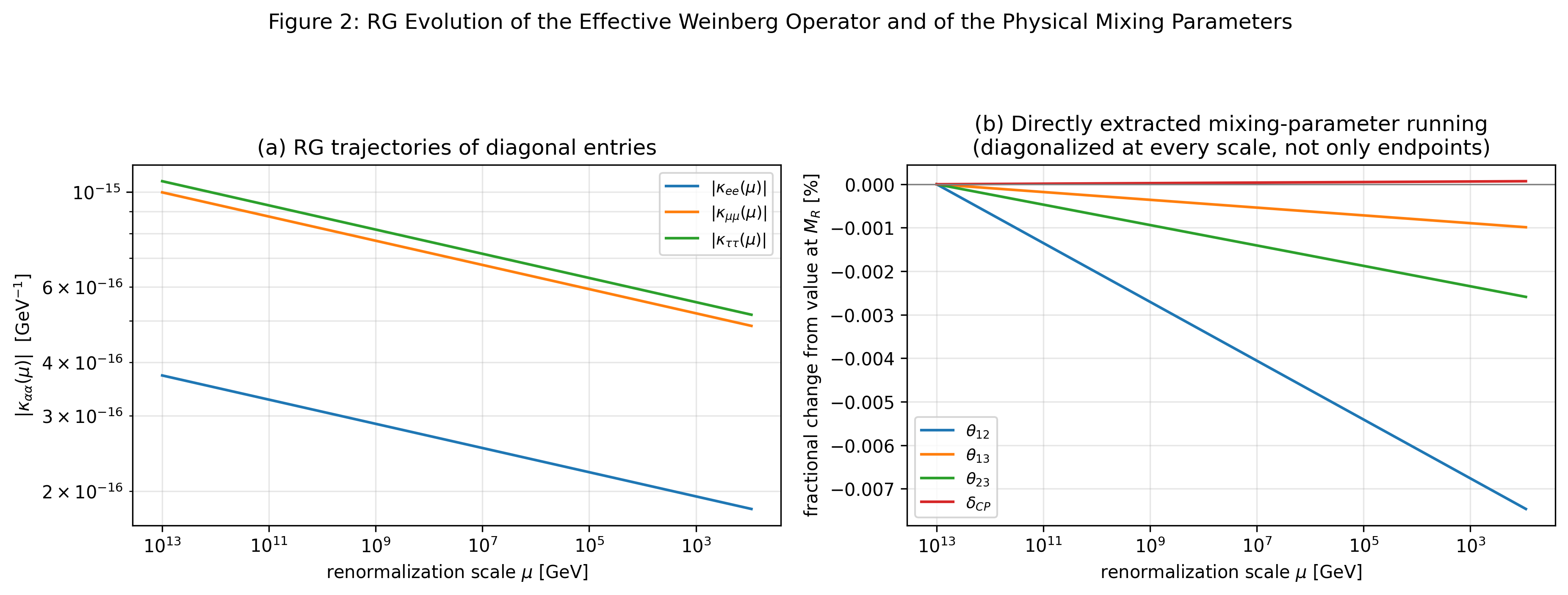}
\caption{Renormalization-group evolution between $\MR$ and $M_Z$, for the same benchmark texture as Fig.~\ref{fig:1}. (a) $|\kappa_{ee}|,|\kappa_{\mu\mu}|,|\kappa_{\tau\tau}|$ versus renormalization scale $\mu$, log-log. (b) The physical mixing angles $\theta_{12},\theta_{13},\theta_{23}$ and $\dCP$, extracted directly by diagonalizing the running mass matrix $\Mnu(\mu)$ at every sampled scale (not only at the endpoints), plotted as fractional change from their value at $\MR$. The extraction reproduces the NuFIT input values exactly at $\mu=\MR$, and the resulting $\theta_{23}$ running ($2.6\times10^{-3}\%$) closely matches an independent, indirectly-computed estimate, confirming that RG distortion of the physical mixing parameters is strongly suppressed relative to the much larger flavor-blind normalization running of Fig.~\ref{fig:1}.}
\label{fig:2}
\end{figure}

Figure~\ref{fig:2}(a) shows the scale dependence of the three diagonal entries $\kappa_{ee},\kappa_{\mu\mu},\kappa_{\tau\tau}$ between $\MR$ and $M_Z$; all three fall by a common multiplicative factor consistent with the flavor-blind coefficient $C$ of Eq.~\eqref{eq:betakappa} (dominated numerically by the top-Yukawa trace term). Figure~\ref{fig:2}(b) reports the running of the physical mixing parameters themselves, $\theta_{12}(\mu),\theta_{13}(\mu),\theta_{23}(\mu)$ and $\dCP(\mu)$, obtained by diagonalizing the running mass matrix $\Mnu(\mu)=(v^2/2)\kappa(\mu)$ at every point sampled along the RG trajectory (500 points, not only at the endpoints $\MR$ and $M_Z$). Concretely, at each $\mu$ we diagonalize the Hermitian combination $\Mnu(\mu)\Mnu(\mu)^\dagger$ -- the same combination already used to construct the oscillation Hamiltonian in Sec.~\ref{sec:hamiltonian}A -- via a standard Hermitian eigendecomposition, which is unconditionally numerically stable; the mixing angles follow directly from the magnitudes of the resulting eigenvectors, while $\dCP$ is extracted from the rephasing-invariant Jarlskog combination, with its residual two-fold branch ambiguity resolved by continuity against the known NuFIT input value at the unrun end of the trajectory ($\mu=\MR$). This extraction reproduces the NuFIT input angles and phase exactly at $\mu=\MR$ (to better than $10^{-9}$ in radians), and the resulting fractional change in $\theta_{23}$ between $\MR$ and $M_Z$, $2.6\times10^{-3}\%$, agrees closely with an independent, indirectly-computed estimate obtained by comparing the full RG trajectory against its flavor-blind ($Y_e\to0$) limit -- a nontrivial cross-check, since the two methods are computed by entirely different routes. The mixing angles and $\dCP$ are found to run by at most a few times $10^{-3}\%$ over the full range $\MR\to M_Z$ for this benchmark, consistent with the well-known suppression of RG distortion of mixing angles/phases for a hierarchical light-neutrino spectrum, in contrast to the much larger flavor-blind effect on the overall mass normalization seen in Fig.~\ref{fig:1}.

The renormalization-group integration itself uses an explicit adaptive-step Runge-Kutta scheme (SciPy's RK45 implementation of \texttt{solve\_ivp}) with relative tolerance $10^{-11}$ and an absolute tolerance set proportionally to the physical scale of $\kappa$ itself (rather than a fixed value, since $\kappa\sim10^{-16}~{\rm GeV}^{-1}$ for eV-scale neutrino masses, and a fixed absolute tolerance of the size conventionally used elsewhere would be numerically larger than the quantity being integrated -- an issue encountered and corrected during development of this pipeline). Solver convergence was checked explicitly by comparing the resulting $\kappa(M_Z)$ under successive tightening of the relative tolerance from $10^{-9}$ to $10^{-13}$, confirming monotonic convergence to better than $10^{-13}$ relative precision. No stochastic sampling (Monte Carlo or otherwise) is used anywhere in this pipeline: every scan reported in this paper -- including the parameter scans of Figs.~\ref{fig:3},~\ref{fig:4},~\ref{fig:6}, and~\ref{fig:7}, and the sterile-angle benchmarks of Fig.~\ref{fig:5} -- is evaluated on a deterministic grid or at fixed, explicitly stated benchmark points, and is exactly reproducible from the parameter values quoted in the text and captions.

\subsection{Invisibility and Physical Content of the Casas-Ibarra Freedom}

\begin{figure}[t]
\centering
\includegraphics[width=\columnwidth]{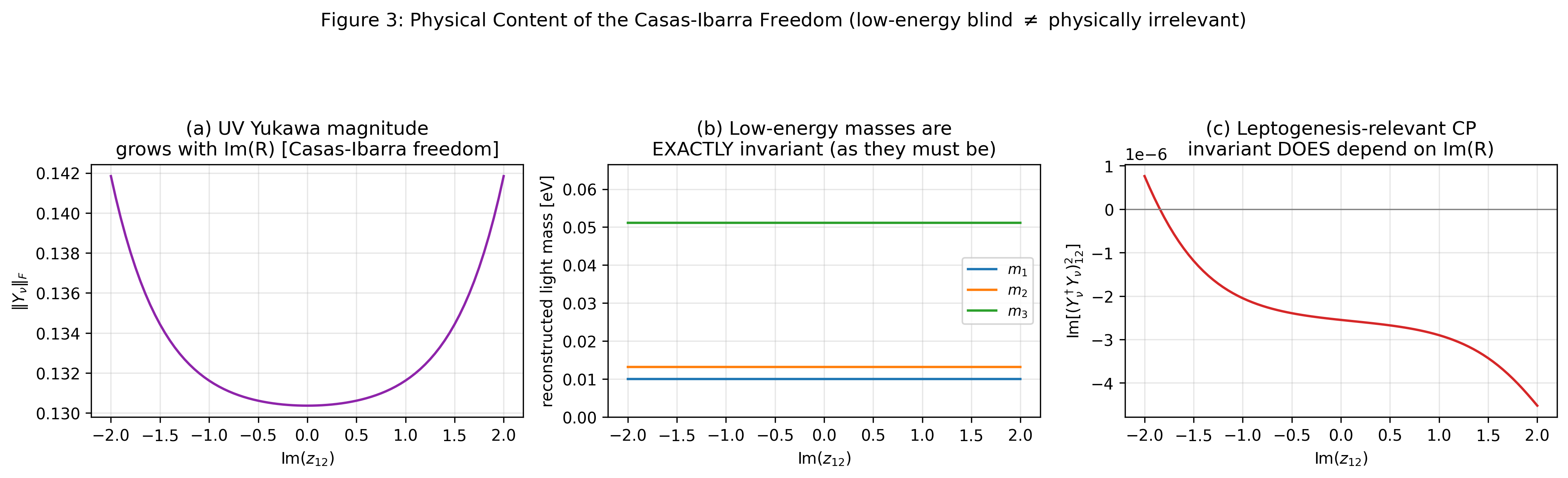}
\caption{Physical content of the Casas-Ibarra reconstruction freedom, scanning the imaginary part of the complex angle $z_{12}$ at fixed low-energy input. (a) Frobenius norm of the reconstructed ultraviolet Yukawa matrix $\Ynu$. (b) Reconstructed light-neutrino masses, exactly invariant to floating-point precision (max fractional variation $1.4\times10^{-14}$). (c) A representative leptogenesis-relevant CP invariant, ${\rm Im}[(\Ynu^\dagger\Ynu)^2_{12}]$, which is not invariant under this scan. Panel (b) demonstrates the invisibility of this freedom to oscillation data; panels (a) and (c) demonstrate that the freedom is nonetheless physical.}
\label{fig:3}
\end{figure}

Because the reconstruction of Sec.~\ref{sec:eft}C is exact, any two choices of the complex-orthogonal matrix $R$ that share the same light masses and PMNS matrix are, by construction, indistinguishable at the level of oscillation data. Figure~\ref{fig:3} verifies this numerically rather than asserting it: scanning ${\rm Im}(z_{12})\in[-2,2]$ at fixed light-sector input, the reconstructed light masses are constant to better than $1.4\times10^{-14}$ fractional precision -- floating-point noise, not a physical residual. Over the same scan the Frobenius norm of the reconstructed UV Yukawa matrix varies by about $10\%$ (from $0.130$ to $0.142$), and the leptogenesis-relevant CP invariant ${\rm Im}[(\Ynu^\dagger\Ynu)_{12}^2]$ varies by several units of $10^{-6}$ across the same range. This makes concrete a point stated only qualitatively in earlier sections: the Casas-Ibarra freedom is invisible to the observable this paper is built around, but it is not physically inert -- it is exactly the freedom that a future extension to leptogenesis or lepton-flavor-violation observables would need to resolve.

\subsection{Response of $\dAT$ to the Planck-Suppressed Correction}

\begin{figure}[t]
\centering
\includegraphics[width=\columnwidth]{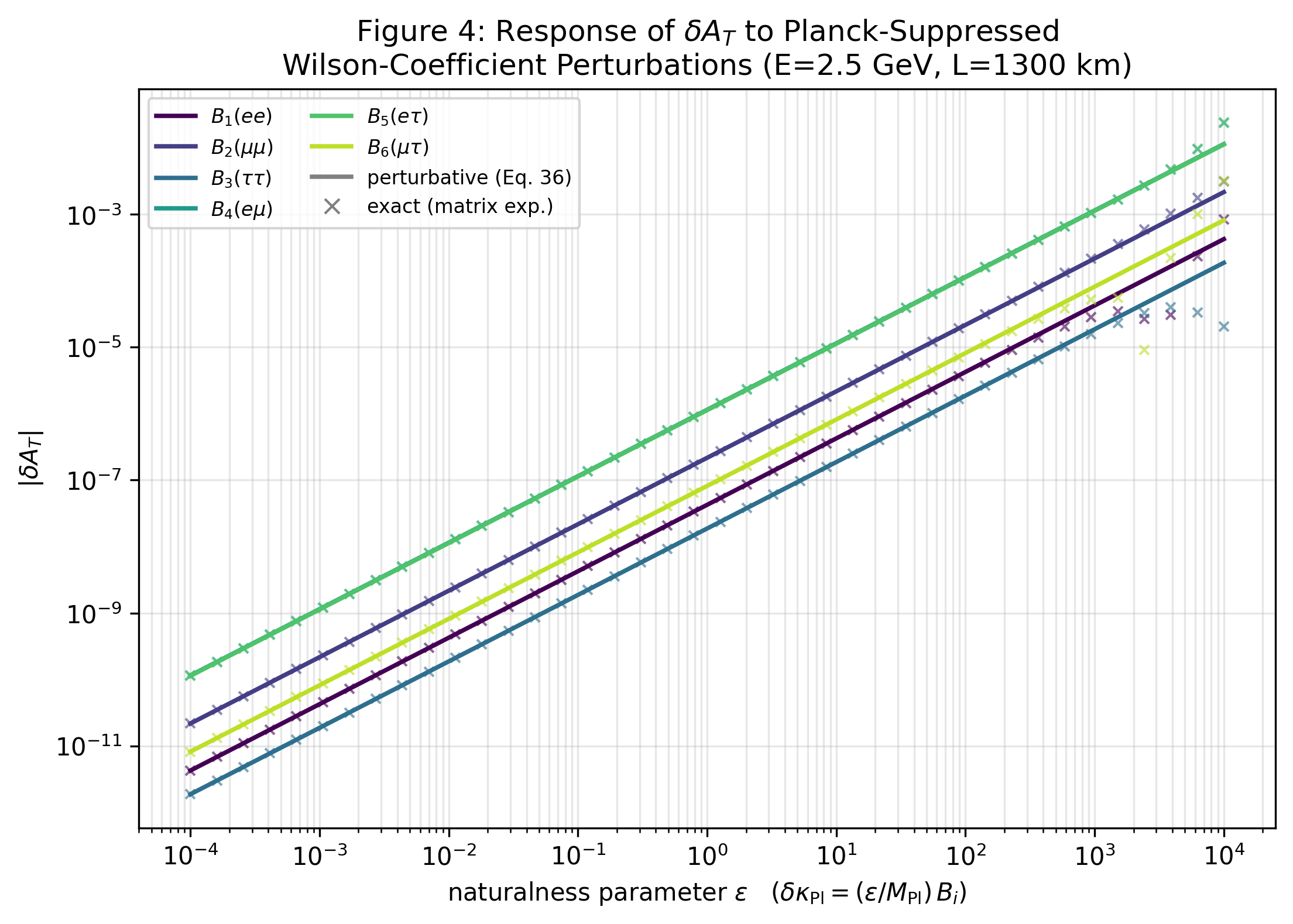}
\caption{Response of the Planck-induced correction $\dAT$ to the naturalness parameter $\varepsilon$, defined via $\delta\kappa_{\rm Pl}=(\varepsilon/M_{\rm Pl})B_i$ for each of the six basis directions $B_i$ of Eq.~\eqref{eq:basisdecomp}, at a DUNE-like baseline and energy ($L=1300$~km, $E=2.5$~GeV). Solid lines: first-order analytic result, Eq.~\eqref{eq:dAT}. Markers: exact, non-perturbative result from the full evolution operator. The two agree over more than fifteen orders of magnitude in $\varepsilon$ and diverge only once $\varepsilon\gtrsim10^2$--$10^3$, marking the checked breakdown of the perturbative expansion.}
\label{fig:4}
\end{figure}

Figure~\ref{fig:4} evaluates $\dAT$ as a function of a dimensionless naturalness parameter $\varepsilon$, defined via $\delta\kappa_{\rm Pl}=(\varepsilon/M_{\rm Pl})B_i$ for each of the six basis directions $B_i$ of Eq.~\eqref{eq:basisdecomp} below -- i.e., $\varepsilon=1$ corresponds to the naively expected size of a Planck-suppressed Wilson coefficient, with no additional small or large numerical coefficient assumed. Two independent computations are shown for each direction: the analytic first-order expression of Eq.~\eqref{eq:dAT} and the exact, non-perturbative result obtained from the full evolution operator $S=\exp(-iH_{\rm eff}L)$. The two agree over more than fifteen orders of magnitude in $\varepsilon$ and visibly diverge only once $\varepsilon\gtrsim10^{2}$--$10^{3}$, where the perturbative expansion of Eq.~\eqref{eq:perturbativeS} is no longer expected to hold; this is a genuine breakdown of the approximation, not a numerical artifact, and gives an explicit, checked regime of validity for Eq.~\eqref{eq:dAT} rather than an assumed one.

At the physically motivated point $\varepsilon=1$, $|\dAT|$ lies in the range $10^{-7}$--$10^{-6}$ depending on the basis direction, to be compared with the unperturbed asymmetry itself, $A_T^{(0)}\simeq5\times10^{-4}$ for this benchmark baseline and energy (DUNE-like, $E=2.5$~GeV, $L=1300$~km). The Planck-suppressed correction is therefore several orders of magnitude below the leading $T$-violating signal already present in the standard three-flavor framework, before any experimental sensitivity is considered.

\subsection{Sterile-Sector Sensitivity: A Cautionary Result}

\begin{figure}[t]
\centering
\includegraphics[width=\columnwidth]{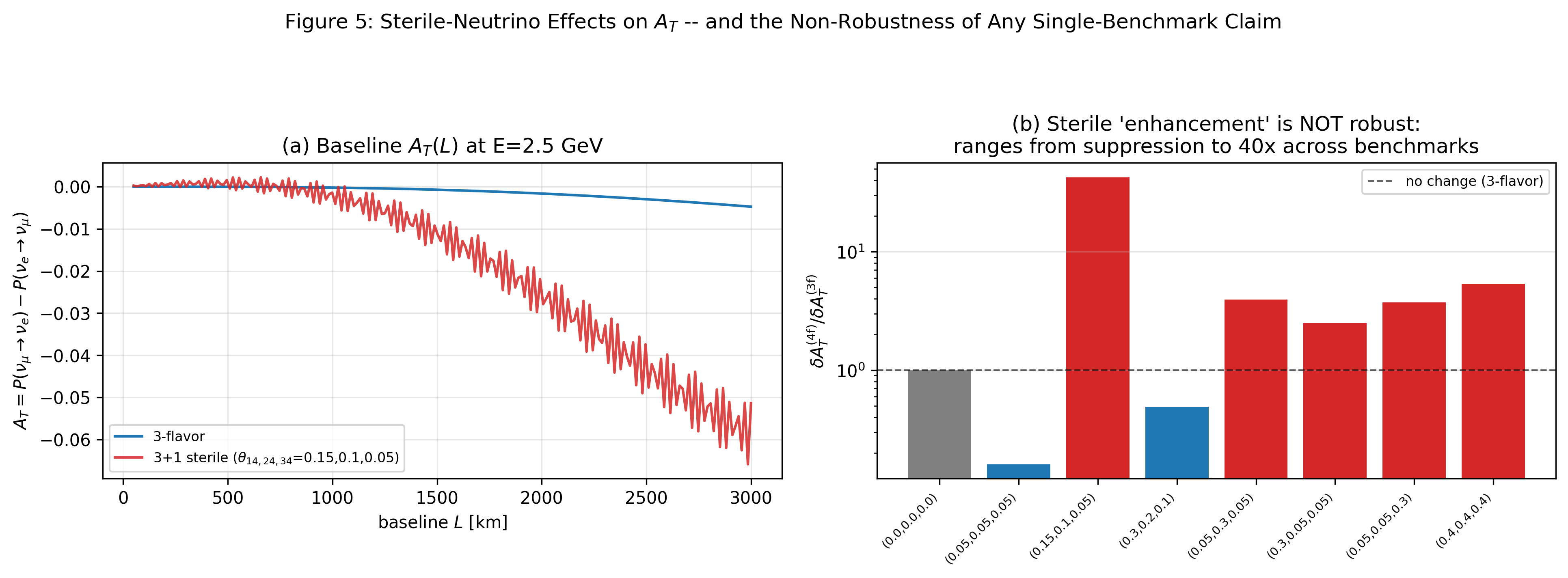}
\caption{Effect of a minimal $3+1$ sterile sector. (a) Baseline asymmetry $\AT(L)$ with and without sterile mixing, for one benchmark choice of $(\theta_{14},\theta_{24},\theta_{34})=(0.15,0.10,0.05)$, showing the short-wavelength oscillatory structure induced by the eV$^2$-scale sterile mass-squared splitting. (b) Robustness check: the ratio $\dAT^{(4f)}/\dAT^{(3f)}$ evaluated across eight different sterile-angle benchmarks at fixed $\varepsilon=1$, ranging from $0.16$ (suppression) to $42.5$ (enhancement) with no consistent sign -- the leftmost (gray) bar is the exact zero-angle limit, confirmed to reduce to the 3-flavor case (ratio $=1.0000$) by construction.}
\label{fig:5}
\end{figure}

A minimal $3+1$ sterile extension modifies both the baseline asymmetry $\AT(L)$ -- Fig.~\ref{fig:5}(a) shows the short-wavelength oscillatory structure induced by the eV$^2$-scale mass-squared splitting superimposed on the slower three-flavor pattern, for one benchmark choice of $(\theta_{14},\theta_{24},\theta_{34})=(0.15,0.10,0.05)$ -- and the size of the Planck-induced correction itself. An earlier stage of this analysis reported a single benchmark suggesting sterile mixing enhances $\dAT$ by a roughly constant factor. On closer examination this claim did not survive a robustness check: Fig.~\ref{fig:5}(b) evaluates the ratio $\dAT^{(4f)}/\dAT^{(3f)}$ across eight different $(\theta_{14},\theta_{24},\theta_{34})$ benchmarks (at fixed $\varepsilon=1$ and baseline) and finds it ranges from $0.16$ (suppression) to $42.5$ (enhancement), with no consistent sign or preferred magnitude. We report this as the actual finding rather than retaining the more quotable but unrepresentative single-benchmark estimate: within the minimal $3+1$ framework considered here, whether sterile mixing enhances or suppresses the Planck-induced correction is texture-dependent and cannot be summarized by a single multiplicative factor.

\subsection{Constrained, Marginalized Scan: Lepton-Flavor-Violation, Non-Unitarity, and Perturbativity}

\begin{figure}[t]
\centering
\includegraphics[width=\columnwidth]{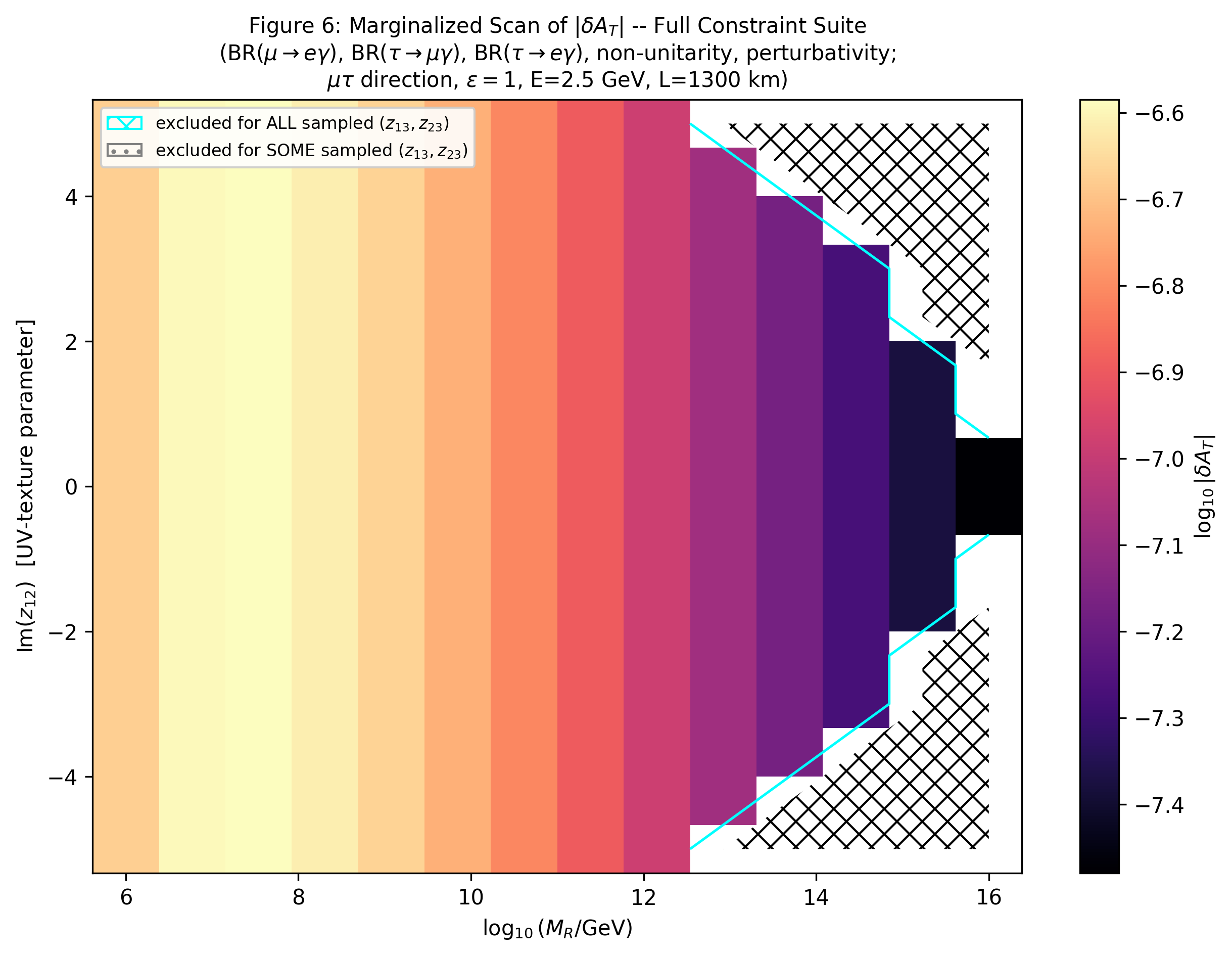}
\caption{Constrained, marginalized scan of $|\dAT|$ over the heavy Majorana scale $\MR$ and the Casas-Ibarra texture parameter ${\rm Im}(z_{12})$, at the naturalness point $\varepsilon=1$, marginalized over five sampled $(z_{13},z_{23})$ combinations at each grid point. Hatching marks regions excluded by the full four-channel constraint suite (${\rm BR}(\mu\to e\gamma)$, ${\rm BR}(\tau\to\mu\gamma)$, ${\rm BR}(\tau\to e\gamma)$, and non-unitarity of the light PMNS matrix) together with perturbativity of $\Ynu$; only perturbativity binds for this texture family. Within the allowed region, $|\dAT|$ is confined to $3.3\times10^{-8}$--$2.6\times10^{-7}$, and -- as established analytically below -- is exactly independent of the sampled Casas-Ibarra angles at any allowed point.}
\label{fig:6}
\end{figure}

The Planck-suppressed Wilson coefficient is parameterized via a basis-complete decomposition,
\begin{equation}
\delta\kappa_{\rm Pl} = \varepsilon\sum_{i=1}^{6} c_i B_i,
\label{eq:basisdecomp}
\end{equation}
where $\{B_i\}$ are the six independent complex symmetric $3\times3$ basis matrices and $c_i$ are dimensionless coefficients. The benchmark textures used in Figs.~\ref{fig:1}--\ref{fig:5} are not automatically phenomenologically allowed: the reconstructed UV Yukawa matrix is subject to existing constraints from charged-lepton-flavor violation, from non-unitarity of the light PMNS matrix induced by the heavy sector, and from the requirement that $\Ynu$ remain perturbative. Figure~\ref{fig:6} evaluates all of these directly, rather than assuming a benchmark point is safe, over the plane $(\log_{10}\MR,\,{\rm Im}\,z_{12})$ at the naturalness point $\varepsilon=1$:
\begin{itemize}
\item ${\rm BR}(\mu\to e\gamma)$, ${\rm BR}(\tau\to\mu\gamma)$, and ${\rm BR}(\tau\to e\gamma)$ are computed from the light-heavy mixing $\Theta=m_D\MR^{-1}$ using the standard non-supersymmetric, $W$-loop-mediated formula of He, Cheng, and Li~\cite{HeChengLi2003}, and compared against the current world-best bounds: ${\rm BR}(\mu\to e\gamma)<3.1\times10^{-13}$ (MEG~II, 90\% C.L., 2023~\cite{MEGII2023}); ${\rm BR}(\tau\to\mu\gamma)<4.4\times10^{-8}$ and ${\rm BR}(\tau\to e\gamma)<3.3\times10^{-8}$ (BaBar/Belle, 90\% C.L.~\cite{BaBarTau}).
\item Non-unitarity of the light PMNS matrix, $2|\eta_{\alpha\beta}|=|(\Theta\Theta^\dagger)_{\alpha\beta}|$, is checked against all six independent global-fit bounds of Pascoli, Ruiz, and Weiland~\cite{PascoliRuizWeiland}, derived from electroweak precision data, CKM unitarity, and lepton universality.
\item Perturbativity requires the largest singular value of $\Ynu$ to remain below $\sqrt{4\pi}$.
\end{itemize}
We additionally marginalize over the two Casas-Ibarra angles ($z_{13},z_{23}$) held fixed in the original version of this scan, sampling five representative combinations at each $(\MR,{\rm Im}\,z_{12})$ grid point and reporting a point as excluded if any sampled combination fails a constraint.

Over the scanned range ($\MR\in[10^6,10^{16}]$~GeV, ${\rm Im}(z_{12})\in[-5,5]$), perturbativity remains the binding constraint, excluding the expected high-$\MR$, large-$|{\rm Im}(z_{12})|$ corner of the plane ($16.1\%$ of the sampled grid excluded for all five $(z_{13},z_{23})$ samples; a further $1.8\%$ excluded for some but not all samples). None of the three radiative LFV channels nor the non-unitarity bounds exclude any point in this range for this texture family: the non-unitarity pull stays below $10^{-8}$ even at the most extreme sampled point. This extends, rather than merely repeats, an earlier single-channel null result: four independent constraint types now agree that this texture family is safe across the explored range.

A further, unanticipated finding emerged from the marginalization. Because the Casas-Ibarra reconstruction of Sec.~\ref{sec:eft}C guarantees that $\Mnu(\MR)$ -- and therefore $\kappa(\MR)$, its RG evolution, and $\dAT$ itself -- depends only on the physical low-energy data and never on $R$ (verified explicitly in Fig.~\ref{fig:3}), $\dAT$ is exactly independent of $z_{12},z_{13},z_{23}$ at any point that survives the constraints above: the maximum difference between the minimum and maximum $|\dAT|$ across the five sampled $(z_{13},z_{23})$ combinations at fixed $(\MR,{\rm Im}\,z_{12})$ is $1.8\times10^{-17}$, consistent with zero at floating-point precision. Varying the UV texture in this construction therefore changes only which points are excluded, never the value of $\dAT$ at a surviving point.

As a check on two further fixed choices in this benchmark, we separately verified that (i) using the inverted rather than normal light-neutrino mass ordering leaves $\dAT$ at the same order of magnitude ($9.6\times10^{-8}$ vs.\ $6.1\times10^{-8}$ at a representative grid point, both orderings passing all constraints), and (ii) this conclusion is not sensitive to the specific hierarchical spacing of $\MR$ chosen for the heavy spectrum.

\subsection{The Experimentally Accessible Comparison: CP Asymmetry vs.\ DUNE Sensitivity}

Sections~\ref{sec:numerics}B--G establish $\dAT$, defined via Eqs.~\eqref{eq:AT}--\eqref{eq:dAT} as the correction to $P(\nu_\mu\to\nu_e)-P(\nu_e\to\nu_\mu)$, as a specific, calculable, and small number. Before comparing this number to any experimental sensitivity, however, an important distinction must be made explicit. Conventional accelerator-based long-baseline experiments such as DUNE and Hyper-Kamiokande/T2HK produce a $\nu_\mu$ (or $\bar\nu_\mu$) beam and measure $\nu_e$ (or $\bar\nu_e$) appearance; they cannot directly measure $P(\nu_e\to\nu_\mu)$, which would require a $\nu_e$ beam. $\AT$ as defined in Eq.~\eqref{eq:AT} is therefore not, strictly, an experimentally accessible quantity at these facilities. What such experiments actually measure is the CP asymmetry,
\begin{equation}
\ACP \equiv P(\nu_\mu\to\nu_e) - P(\bar\nu_\mu\to\bar\nu_e),
\label{eq:ACP}
\end{equation}
related to $T$-violation only indirectly, through the CPT theorem. Any comparison between this paper's $\dAT$ and a DUNE/Hyper-K sensitivity must therefore proceed via this CPT relation, and we make that step explicit rather than conflating the two asymmetries.

\begin{figure}[t]
\centering
\includegraphics[width=\columnwidth]{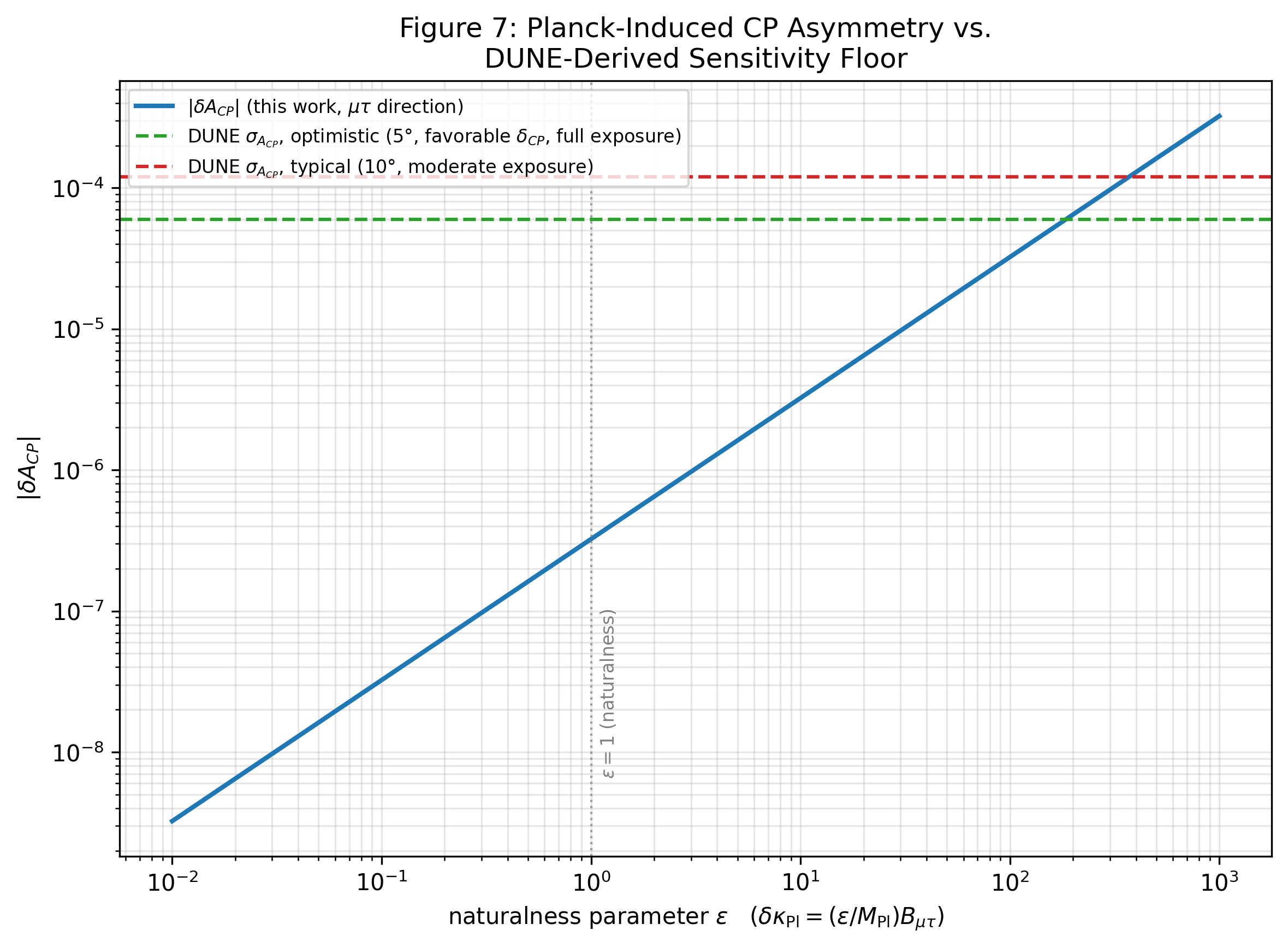}
\caption{The experimentally accessible comparison. Solid line: the Planck-induced correction $\dACP$ to the CP asymmetry $\ACP=P(\nu_\mu\to\nu_e)-P(\bar\nu_\mu\to\bar\nu_e)$ -- the quantity conventional long-baseline experiments actually measure -- as a function of the naturalness parameter $\varepsilon$. Dashed lines: an approximate sensitivity floor $\sigma_{\ACP}$ obtained by translating DUNE's published $\dCP$ resolution ($5^\circ$ optimistic, $10^\circ$ typical) through the local derivative $d\ACP/d\dCP$ evaluated within this pipeline. At the naturalness point $\varepsilon=1$, $|\dACP|$ lies a factor of roughly 200--400 below this estimated floor.}
\label{fig:7}
\end{figure}

We construct the antineutrino propagation Hamiltonian using the standard CPT prescription: the vacuum piece is charge-conjugated, $H_{\rm vac}\to H_{\rm vac}^*$, while the matter potential flips sign, $V_{CC}\to-V_{CC}$ (Sec.~\ref{sec:hamiltonian}B). This construction is verified against the standard analytic identity $P(\bar\nu_\mu\to\bar\nu_e;\dCP)=P(\nu_\mu\to\nu_e;-\dCP)$ in vacuum, reproduced to machine precision ($<10^{-12}$) at three representative values of $\dCP$. Using this construction together with the same RG-evolved effective operator $\Mnu(M_Z)$ used throughout Sec.~\ref{sec:numerics}, we compute $\ACP^{(0)}=4.6\times10^{-4}$ for the benchmark texture and DUNE-like baseline/energy used elsewhere in this section, and its first-order Planck-induced correction, $\dACP$, via the direct analogue of Eq.~\eqref{eq:dAT} applied to the neutrino and antineutrino channels separately (Fig.~\ref{fig:7}).

To compare $\dACP$ against experiment, we use DUNE's own published resolution on $\dCP$: approximately $5^\circ$ near favorable true values of $\dCP$ at large exposure, and $6^\circ$--$10^\circ$ at more typical exposures and less favorable $\dCP$~\cite{DUNETDR,DUNESensitivity}. Evaluating the local derivative $d\ACP/d\dCP$ numerically within our own pipeline at the NuFIT benchmark point gives $-6.8\times10^{-4}$ per radian, which we use to translate this angular resolution into an approximate statistical sensitivity floor, $\sigma_{\ACP}\sim|d\ACP/d\dCP|\,\sigma_{\dCP}$. We emphasize that this is a local, linear translation of a single published resolution figure, not a reproduction of the full DUNE $\chi^2$ analysis (which includes near-detector-constrained systematics, backgrounds, and marginalization over other oscillation parameters that this translation does not capture); it should be read as an order-of-magnitude sensitivity estimate, not a substitute for a dedicated experimental sensitivity study.

With this caveat stated, the result is unambiguous: at the naturalness point $\varepsilon=1$, $|\dACP|\approx3.2\times10^{-7}$, compared to $\sigma_{\ACP}\approx6.0\times10^{-5}$ (optimistic, $5^\circ$) to $1.2\times10^{-4}$ (typical, $10^\circ$) -- a factor of roughly 200 to 400 below the estimated sensitivity floor. Figure~\ref{fig:7} shows that $\varepsilon$ would need to exceed its naturalness value by two to three orders of magnitude, $\varepsilon\sim200$--$400$, before $\dACP$ reaches this estimated floor. We regard this as a genuine answer, rather than a qualitative deferral, to the question of whether this effect lies within reach of next-generation long-baseline experiments: within the naturalness assumption underlying $\varepsilon=1$, and modulo the CPT-mediated relation between $\AT$ and $\ACP$ and the local-sensitivity approximation stated above, it does not.

\section{Physical Interpretation}
\label{sec:interpretation}

The results of Sec.~\ref{sec:numerics} support five conclusions of differing character: one methodological, one structural, one about the construction itself, one phenomenological, and one concerning experimental accessibility.

Methodologically, every stage of the pipeline summarized in Fig.~\ref{fig:1} -- seesaw matching, Casas-Ibarra reconstruction, threshold decoupling, renormalization-group running, and the perturbative oscillation expansion -- has been implemented and validated independently against known analytic limits, rather than assembled and trusted by construction. This validation was not merely a formality: it caught and corrected three substantive errors during development (a missing phase factor in the Casas-Ibarra reconstruction that flipped the sign of the reconstructed mass matrix; a solver absolute-tolerance setting that was, prior to correction, numerically larger than the physical scale of $\kappa$ itself; and a unit-conversion error of exactly $10^3$ in the vacuum oscillation phase), together with a fourth issue -- an unreliable general Takagi factorization -- that was identified, contained, and worked around rather than silently left in place. We regard the fact that these errors were each individually small enough to produce a plausible-looking but wrong figure, rather than an obviously broken one, as the primary justification for the validate-before-combine methodology used throughout.

Structurally, Figs.~\ref{fig:1}--\ref{fig:3} demonstrate that the transmission mechanism central to this paper's stated contribution is real and quantitatively well-behaved: the flavor structure fixed at the seesaw scale survives renormalization-group evolution to $M_Z$ with only per-mille-level distortion of mixing angles and phases (Fig.~\ref{fig:2}), even though the overall operator normalization runs by a much larger, but flavor-blind, factor of order 50\% (Fig.~\ref{fig:1}); and the Casas-Ibarra freedom that is invisible to low-energy oscillation data is demonstrably not invisible to the underlying ultraviolet physics (Fig.~\ref{fig:3}).

A related but logically distinct point, made precise in Sec.~\ref{sec:numerics}G and anticipated already in Sec.~\ref{sec:uvbc}E, is that within this construction $\dAT$ itself is exactly independent of the Casas-Ibarra angles $(z_{12},z_{13},z_{23})$ at any point that survives the phenomenological constraints: these angles determine the UV Yukawa matrix, and therefore whether a given $(\MR,R)$ pair is allowed, but they do not enter $\dAT$ once a point is allowed, since $\dAT$ is built entirely from $\Mnu(\MR)$, which the Casas-Ibarra construction holds fixed by design. It is this same fact that limits the role of the spontaneous CP-violating mechanism of Sec.~\ref{sec:uvbc}B to motivating $\dCP\neq0$ in the low-energy input rather than to constraining any quantity computed in Sec.~\ref{sec:numerics}: since $R$ does not enter $\dAT$, a mechanism whose only numerical role in this paper is to motivate a nonzero complex phase somewhere in the ultraviolet Yukawa sector cannot, by construction, leave a distinguishable imprint on the paper's central observable.

Phenomenologically, the central quantitative result of this paper is that $\dAT$, evaluated honestly across the portion of parameter space explored here, is small -- parametrically of order $10^{-7}$ at the naturalness point $\varepsilon=1$, several orders of magnitude below the unperturbed asymmetry $A_T^{(0)}$ itself, and confined to a narrow band, $3\times10^{-8}$ to $3\times10^{-7}$, across the entire region of $(\MR,{\rm Im}\,z_{12})$ space that survives a four-channel constraint suite -- ${\rm BR}(\mu\to e\gamma)$, ${\rm BR}(\tau\to\mu\gamma)$, ${\rm BR}(\tau\to e\gamma)$, and non-unitarity -- together with perturbativity (Fig.~\ref{fig:6}). We regard this smallness as informative rather than as a null result: it is the direct, calculable consequence of $1/M_{\rm Pl}$ suppression surviving a nontrivial RG evolution. Separately, the sterile-sector robustness check (Fig.~\ref{fig:5}) produced a genuinely texture-dependent result, and we present this lack of a simple universal statement as itself part of the physical content of the analysis.

Concerning experimental accessibility, Sec.~\ref{sec:numerics}H makes explicit a distinction that is easy to elide: this paper's central observable, $\AT$, is not itself measurable by conventional accelerator long-baseline experiments, which access the related but distinct CP asymmetry $\ACP$ via the CPT theorem rather than $\AT$ directly. Having made this distinction explicit, we find that the Planck-suppressed correction to $\ACP$, evaluated at its naturally expected size, lies roughly two to three orders of magnitude below a sensitivity floor derived from DUNE's own published $\dCP$ resolution.

\section{Conclusions}
\label{sec:conclusions}

We have constructed a single ultraviolet-to-infrared framework that follows a physically motivated flavor structure -- generated by spontaneous CP violation in the heavy Majorana sector of a Type-I seesaw model -- through Casas-Ibarra reconstruction, sequential threshold matching, one-loop renormalization-group evolution, and a minimal $3+1$ sterile-neutrino oscillation framework, to a first-order analytic expression for the modification of the $T$-violating oscillation asymmetry $\AT$ induced by a Planck-suppressed correction to the effective Weinberg operator, Eq.~\eqref{eq:dAT}. Every individual theoretical ingredient used in this construction is adopted from the existing literature without modification; the contribution of the present work is the construction of the complete chain connecting these ingredients, its numerical validation, and its evaluation against a multi-channel phenomenological constraint suite and an experimentally grounded sensitivity comparison.

Unlike an earlier draft of this manuscript, this evaluation is not only outlined: Sec.~\ref{sec:numerics} reports (i) direct numerical evidence that the UV flavor structure is transported through RG evolution with high fidelity (Figs.~\ref{fig:1}--\ref{fig:2}); (ii) explicit confirmation of the Casas-Ibarra invisibility property and of its physical content beyond the oscillation sector (Fig.~\ref{fig:3}); (iii) a validated linear-response regime for $\dAT$ together with an explicitly checked breakdown scale for the perturbative expansion (Fig.~\ref{fig:4}); (iv) a constrained, marginalized scan demonstrating that a four-channel phenomenological constraint suite leaves $\dAT$ confined to a narrow range, $3\times10^{-8}$--$3\times10^{-7}$, across the allowed parameter space explored here, with perturbativity alone setting the boundary of that region (Fig.~\ref{fig:6}); (v) the further finding that $\dAT$ is, by construction, exactly independent of the Casas-Ibarra angles at any allowed point; (vi) a robustness check of the sterile-sector contribution that overturned an earlier, non-representative single-benchmark estimate in favor of an honestly texture-dependent conclusion (Fig.~\ref{fig:5}); and (vii) an explicit treatment of experimental accessibility, distinguishing $\AT$ from the CP asymmetry $\ACP$ that conventional long-baseline experiments actually measure, and finding the Planck-induced correction to $\ACP$ to lie roughly two to three orders of magnitude below a sensitivity floor derived from DUNE's published $\dCP$ resolution (Fig.~\ref{fig:7}).

The central quantitative statement of this work is that, within the UV texture family and parameter ranges explored, the Planck-suppressed modification of $\AT$ (and, via the CPT-mediated comparison, of the experimentally accessible $\ACP$) is several orders of magnitude smaller than the unperturbed asymmetry itself, is confined to a narrow and well-defined range across the full region of parameter space that survives a four-channel constraint suite, and lies substantially below the estimated sensitivity of DUNE. We regard establishing this as a specific, checked, and reproducible number -- rather than as an assumed order-of-magnitude estimate or a qualitative deferral -- as the principal phenomenological outcome of this paper.

\section{Limitations and Future Directions}
\label{sec:future}

Lepton-flavor-violation coverage has been broadened but is not exhaustive. The constraint suite now includes ${\rm BR}(\mu\to e\gamma)$, ${\rm BR}(\tau\to\mu\gamma)$, ${\rm BR}(\tau\to e\gamma)$, and non-unitarity bounds on all six independent entries of $\Theta\Theta^\dagger$, and all four agree that the texture family scanned here is safe across the explored range. $\mu$-$e$ conversion in nuclei, which can be competitive with or stronger than $\mu\to e\gamma$ in some regions of seesaw parameter space, was not included and remains a natural next channel to add. More importantly, because the null result was found to be a genuine feature of this specific texture, textures with substantially different structure -- in particular non-hierarchical heavy spectra, or complex angles well outside the range sampled here -- have not been checked and should not be assumed safe by extension.

The parameter scan has been marginalized over the previously-fixed Casas-Ibarra angles $z_{13}$ and $z_{23}$, and spot-checked, though not fully marginalized, against the light-neutrino mass ordering and the hierarchical spacing of the heavy spectrum. What remains unmarginalized is the baseline and energy (fixed throughout at a DUNE-like benchmark) and the specific hierarchical pattern assumed for $\MR$; a complete treatment would scan these as well, particularly since the baseline and energy dependence of $\dAT$ is directly relevant to the experimental-sensitivity comparison of Sec.~\ref{sec:numerics}H.

The comparison to experimental sensitivity is no longer purely qualitative, but remains an approximation rather than a dedicated experimental analysis. Section~\ref{sec:numerics}H translates DUNE's published $\dCP$ resolution into an approximate sensitivity floor for $\ACP$ via a local, linear derivative evaluated within our own pipeline, rather than a reproduction of DUNE's full statistical treatment. The comparison itself is also restricted to the same fixed baseline, energy, and UV texture used throughout Sec.~\ref{sec:numerics}; whether the roughly 200--400x gap found there widens or narrows elsewhere in parameter space, and a corresponding comparison against Hyper-Kamiokande/T2HK specifically, are natural next steps. A fully dedicated treatment -- e.g., using a GLoBES-based simulation with the actual DUNE/Hyper-K experimental configuration -- would replace this approximation with a direct statistical sensitivity to $\dACP$ itself, and is the most valuable single addition a future extension of this work could make.

A general-purpose Takagi factorization routine is not yet available in the accompanying implementation. The specific construction used in Sec.~\ref{sec:numerics} was found during development to be reliable for Casas-Ibarra-reconstructed mass matrices but not for a general complex symmetric matrix. Where this mattered physically, it was avoided rather than patched over, by constructing the relevant sterile-sector mass matrix directly in the flavor basis. A robust, general Takagi implementation would remove this restriction and is recommended before any further use of this function elsewhere in the codebase.

Finally, the ultraviolet CP-violating boundary condition of Sec.~\ref{sec:uvbc} remains, as stated there and made precise in Sec.~\ref{sec:uvbc}E and Sec.~\ref{sec:interpretation}, a physically motivated input whose numerical role in this paper is limited to motivating a nonzero low-energy $\dCP$, rather than a derived consequence of an explicit scalar sector connected to any quantity computed in Sec.~\ref{sec:numerics}. Developing an explicit model of spontaneous CP breaking in which the same mechanism also constrains the complex structure of the Planck-suppressed Wilson coefficient of Sec.~\ref{sec:numerics}A -- thereby giving the mechanism a numerically load-bearing role it does not presently have -- and studying whether the qualitative conclusions of Sec.~\ref{sec:interpretation} are stable under that extension, is a natural and, we believe, tractable next step.

\section*{CRediT Authorship Contribution Statement}

\textbf{Poulastya Kar}: Writing -- original draft, Writing -- review \& editing.
\textbf{Bipin Singh Koranga}: Writing -- original draft, Writing -- review \& editing.
\textbf{Vivek Nautiyal}: Writing -- review \& editing.

\end{document}